\documentclass[aps,prb,twocolumn,superscriptaddress,groupedaddress,footinbib]{revtex4-1} 

\usepackage{longtable}
\usepackage{morefloats}
\usepackage[dvips]{graphicx}
\usepackage{color}
\usepackage{epsfig,graphicx,amsfonts,amsbsy}
\usepackage{amsmath,amsfonts,amsthm,amssymb}
\usepackage{appendix}
\usepackage{makeidx}
\usepackage{url}
\usepackage{verbatim}
\usepackage[bookmarksnumbered,pdfpagelabels=true,plainpages=false,colorlinks=true,linkcolor=blue,citecolor=red,urlcolor=blue]{hyperref}
\usepackage[rightcaption]{sidecap}
\usepackage{array}
\usepackage{multirow}
\usepackage{tabularx}
\usepackage{braket} 
\usepackage{dsfont}
\usepackage{bbold}
\usepackage{etoolbox}
\usepackage{comment}

\usepackage{booktabs}
\usepackage{siunitx}
\usepackage{multirow}
\newcolumntype{P}[1]{>{\centering\arraybackslash}p{#1}}
\newcolumntype{M}[1]{>{\centering\arraybackslash}m{#1}}

\begin{document}

\title{Decoherence in Andreev spin qubits}
\author{Silas Hoffman,$^{1}$ Max Hays,$^{2}$ Kyle Serniak$^{2,3}$, Thomas Hazard,$^{3}$ Charles Tahan$^{1}$}
\affiliation{$^1$Laboratory for Physical Sciences, 8050 Greenmead Drive, College Park, Maryland 20740, USA}
\affiliation{$^2$Research Laboratory of Electronics, Massachusetts Institute of Technology, Cambridge, MA 02139, USA}
\affiliation{$^3$Lincoln Laboratory, Massachusetts Institute of Technology, 244 Wood Street, Lexington, Massachusetts 02421, USA}

\date{\today}
	
\begin{abstract}
We theoretically study the dephasing of an Andreev spin qubit (ASQ) due to electric and magnetic noise. Using a tight-binding model, we calculate the Andreev states formed in a Josephson junction where the link is a semiconductor with strong spin-orbit interaction. As a result of both the spin-orbit interaction and induced superconductivity, the local charge and spin of these states varies as a function of externally controllable parameters: the phase difference between the superconducting leads, an applied magnetic field, and filling of the underlying semiconductor. Concomitantly, coupling to fluctuations of the electric or magnetic environment will vary, which informs the rate of dephasing. We qualitatively predict the dependence of dephasing on the nature of the environment, magnetic field, phase difference between the junction, and filling of the semiconductor. Comparing the simulated electric- and magnetic-noise-induced dephasing rate to experiment suggests that the dominant source of noise is magnetic. Moreover, by appropriately tuning these external parameters, we find sweet-spots at which we predict an enhancement in ASQ coherence times.
\end{abstract}

\maketitle

\section{Introduction}

Andreev bound states (ABSs) are electron-hole composite states which are localized between two superconducting leads\cite{kulik1969macroscopic, beenakker1991june, Furusaki1991}. While ABSs are typically spin-degenerate, an applied magnetic field and/or spin-orbit interaction combined with a phase bias can break that spin-degeneracy\cite{beriSplitting2008, reynoso2012spin, yokoyama2014anomalous, cayaoSNS2015, muraniAndreev2017, vanHeckZeeman2017, parkPRB17, tosi2019spin, hays2020continuous, bargerbos2022spectroscopy}.
In this regime, the quasiparticle population of two spinful Andreev levels can be controlled and used to encode quantum information.
Such a two level system has been so-named an Andreev spin qubit (ASQ)~\cite{chtchelkatchev2003andreev, padurariu2010theoretical, hays2021CoherentASQ, pita2023direct}.

ASQs can be viewed as a marriage of semiconducting spin qubits and superconducting qubits~\cite{hanson2007spins, krantz2019quantum}. Consequently, they react to external controls typically used in both spin and superconducting qubits. For instance, an applied magnetic field can control the spectrum by both a Zeeman splitting and flux through the superconducting circuit; the qubit can be manipulated using electric dipole spin resonance \cite{pita2023direct, metzger2020} or using microwave currents \cite{hays2021CoherentASQ, cerrillo2021spin, fauvel2023opportunities}.
While rapid control\cite{pita2023direct} and strong coupling\cite{pita2023strong} of ASQs have been demonstrated, measured coherence times are on the order of tens of nanoseconds, orders of magnitude shorter than the corresponding state-of-the-art superconducting and semiconducting spin qubits.

Although several experimental clues have helped to indicate that the dominant contributor to decoherence in ASQs\cite{hays2021CoherentASQ, pita2023direct} is pure dephasing, no systemic attribution of errors currently exists. In this manuscript, we aim to help fill this gap in the literature by numerically studying the source of pure dephasing in ASQs. In particular, we study realistic ASQ systems generated within a 1D tight-binding model and coupled to a bath of spin or charge fluctuators. We predict a qualitative dependence on the decoherence time as a function of the flux and Zeeman splitting and find that this dependence is largely due to the charge or spin of the states that encode the qubit. This description provides a recipe for determining the primary decoherence source and, thus, potential modes of mitigation. Upon comparing with recent experimental results, we find that the decoherence profile best matches a noisy spin bath. More specifically, a spin bath interacting with the Andreev wavefunctions over a long range. Moreover, we find that appropriately tuning the filling of the underlying semiconductor renormalizes the magnetic moment of the Andreev levels which decreases the effective coupling of the spins to the ASQ and increases the coherence time.

The remainder of this manuscript is organized as follows: in Sec.~\ref{sec:Andreev states} we introduce the tight-binding model used to generate an ASQ spectrum and discuss their charge and spin properties. We discuss the models of noise we use and how they couple to ASQs in Sec.~\ref{sec:Noise Models}. In Sec.~\ref{sec:Decoherence} we calculate the decoherence rates associated with the Andreev states found in Sec.~\ref{sec:Andreev states} as a result of coupling to electric and magnetic baths. Upon generating a realistic Andreev spectrum, matched to experiment, in Sec.~\ref{sec:realistic}, we compare simulated decoherence rates with that experiment~\cite{pita2023direct}. In Sec.~\ref{sec:outlook}, we discuss how our results could be extended and, in Sec.~\ref{sec:summary}, we summarize our results.

\section{Andreev states}
\label{sec:Andreev states}

\subsection{Model Hamiltonian and spectrum}
While analytical models of Andreev spin qubits are available,\cite{parkPRB17} in order to incorporate various noise models, a numerical description of the system is useful. We use a 1D tight-binding model incorporating two coupled bands, superconductivity, and spin-orbit interaction: 
\begin{align}
    H_{TB}&=\sum_{j=1}^{N-1} \left(t_j C_{j}^\dagger \tau_z C_{j+1}+i\alpha C_{j}^\dagger\sigma_z\tau_z C_{j+1}+\textrm{H.c.}\right)\nonumber\\
    &+\sum_{j=1}^{N}\left[\mu C_{j}^\dagger\tau_z C_{j}+ \mu' C_j^\dagger  (\mathbb 1_{2\times2}+\rho_z)\tau_zC_{j} \right.\nonumber\\
    &+\Delta_j\cos\phi_j C_{j}^\dagger \sigma_y\tau_y C_{j}+\Delta_j\sin\phi_j C_{j}^\dagger \sigma_y\tau_x C_{j}\nonumber\\
    &+ \left.\alpha' C_{j}^\dagger \sigma_y\rho_y\tau_z C_{j}+B^z C_j^\dagger\sigma_z\tau_z C_j\right]\,.
    \label{HTB}
\end{align}
Here, $C_j=(c_{j\uparrow1},c_{j\downarrow1},c_{j\uparrow2},c_{j\downarrow2},c_{j\uparrow1}^\dagger,c_{j\downarrow1}^\dagger,c_{j\uparrow2}^\dagger,c_{j\downarrow2}^\dagger)$ and $c_{j\sigma n}$ annhilates an electron at site $j$, with spin $\sigma$, in subband $n$. $\sigma_\beta$, $\tau_\beta$, and $\rho_\beta$ for $\beta=x,y,z$, are Pauli matrices acting on the spin, particle-hole, and band space, respectively. Additionally, $t_j$ is the hopping amplitude, $\mu$ is the chemical potential, $\alpha$ is the spin orbit coupling strength, $\Delta_j$ ($\phi_j$) is the position dependent superconducting pairing strength (phase), $\mu'$ is the energy difference between the subbands, and $\alpha'$ is the mixing between the subbands.\footnote{While $\mu'$ and $\alpha'$ could be generated by a two-dimensional lattice with in-plane spin orbit interaction, it is convenient for control to parameterize them independently. Moreover, given the cross section of a quasi-one-dimensional junction, one can calculate $\mu'$ and $\alpha'$ according to Ref.~\onlinecite{parkPRB17}} $B^z$ is the Zeeman splitting as a result of a magnetic field applied along the spin-orbit polarization direction.

\begin{figure}
\includegraphics[width=\columnwidth]{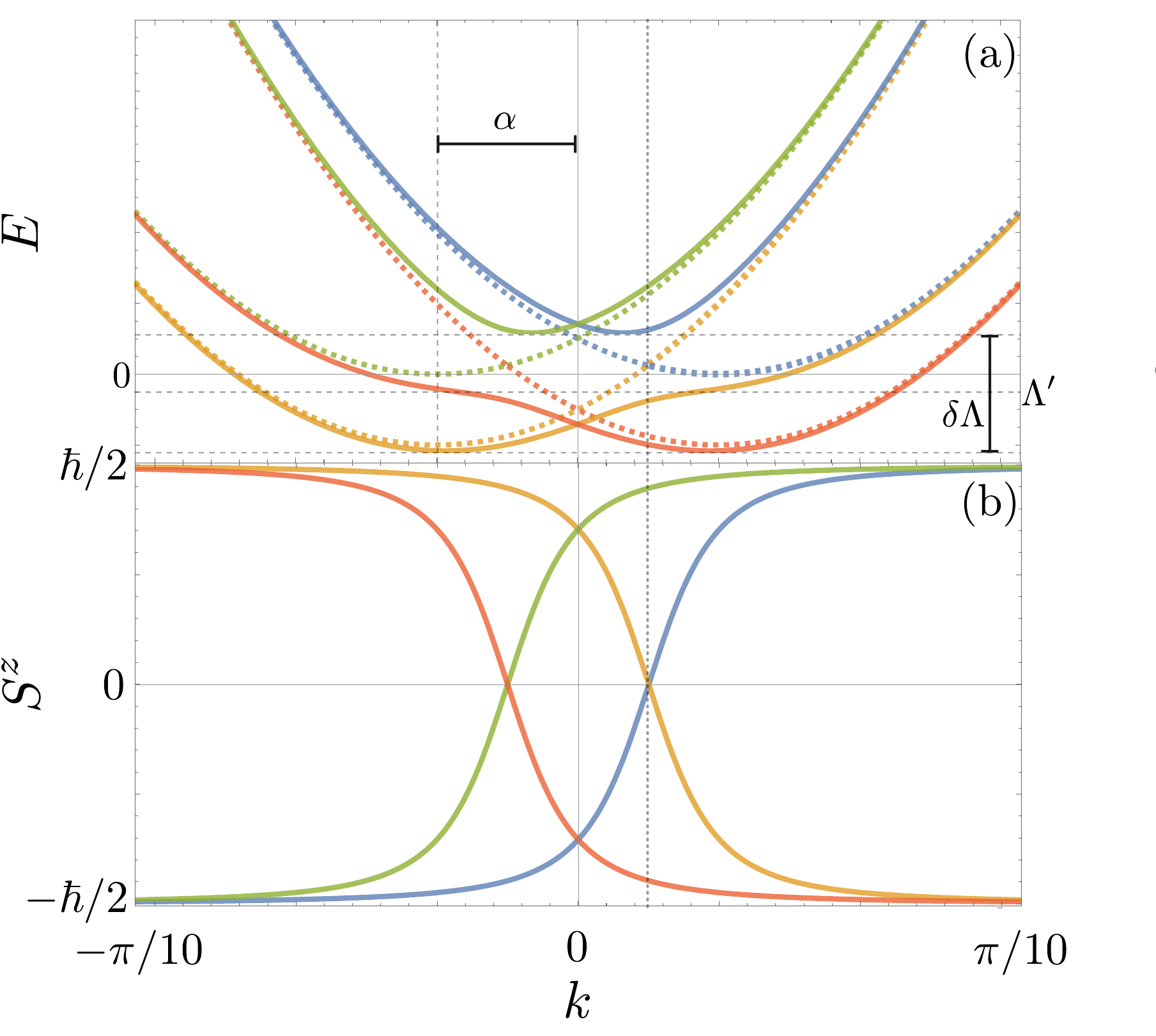}
\caption{Spectrum of the infinite semiconductor in the absence of superconductivity (a), i.e. the eigenvalues of Eq.~(\ref{Hk}), and the corresponding magnetic moment (b) which is a function of the momentum, $k$. The parameters are $10^2\mu'=10(\mu+2t)=10\alpha=t=1$ and $\alpha'=10^{-2}$ (solid) and $\alpha'=0$ (dashed) when the bands are hybridized and uncoupled, respectively. $\Lambda'$ is the energy, with respect to the chemical potential, at which the Fermi velocity is at a local minimum. $\delta\Lambda$ is the energy difference between the bottoms of the hybridized bands (solid curves). The vertical dotted line is the value of $k$ at which the low- and high-energy bands cross or anticross.}
\label{infspec}
\end{figure}

Because many of the properties of the Andreev states will be inherited from the underlying material, it is instructive to examine the low-energy eigenstates of Eq.~(\ref{HTB}) in the absence of superconductivity, $\Delta_j=0$. For a spatially homogeneous infinite system, $t_j=t$, the Fourier transformed Hamiltonian in the momentum basis is
\begin{align}
    H_{TB}&=\sum_k C_k^\dagger \mathcal H_k C_k\,,\nonumber\\
    \mathcal H_k&=(2 t \cos k  + \mu)\tau_z +  2 \alpha \sin k \sigma_z\tau_z + 
 \mu'\tau_z(\mathbb 1_{2\times2}+\rho_z) \nonumber\\
 &+\alpha' \sigma_y\tau_z\rho_y\,,
 \label{Hk}
\end{align}
where $C_k$ is the Fourier transform of $C_j$. In Fig.~\ref{infspec}(a) we plot the spectrum for the continuum, i.e. diagonalizing Eq.~(\ref{Hk}). For a value of the chemical potential wherein only the lower energy band is filled, there are four Fermi points which are characteristic of quasi-one-dimensional spin-orbit systems. In the absence of interband coupling [dashed curves in Fig.~\ref{infspec}(a)], there exists only one Fermi velocity and the spin of the bands are quantized to values of $\hbar/2$. When the subbands are coupled [solid curves in Fig.~\ref{infspec}(a)], the hybridization generates two different Fermi velocities. Moreover, precisely because of this interband coupling, the spin of the eigenstates is momentum-dependent. When $|k|\gg0$, the eigenstates are spin-1/2 states. However, because the two hybridized states have opposite spin, the spin of the eigenstates is generally reduced, e.g. the spin of the eigenstates is zero near the anticrossing [solid curves in Fig.~\ref{infspec}(b)].

\begin{figure}
\includegraphics[width=\columnwidth]{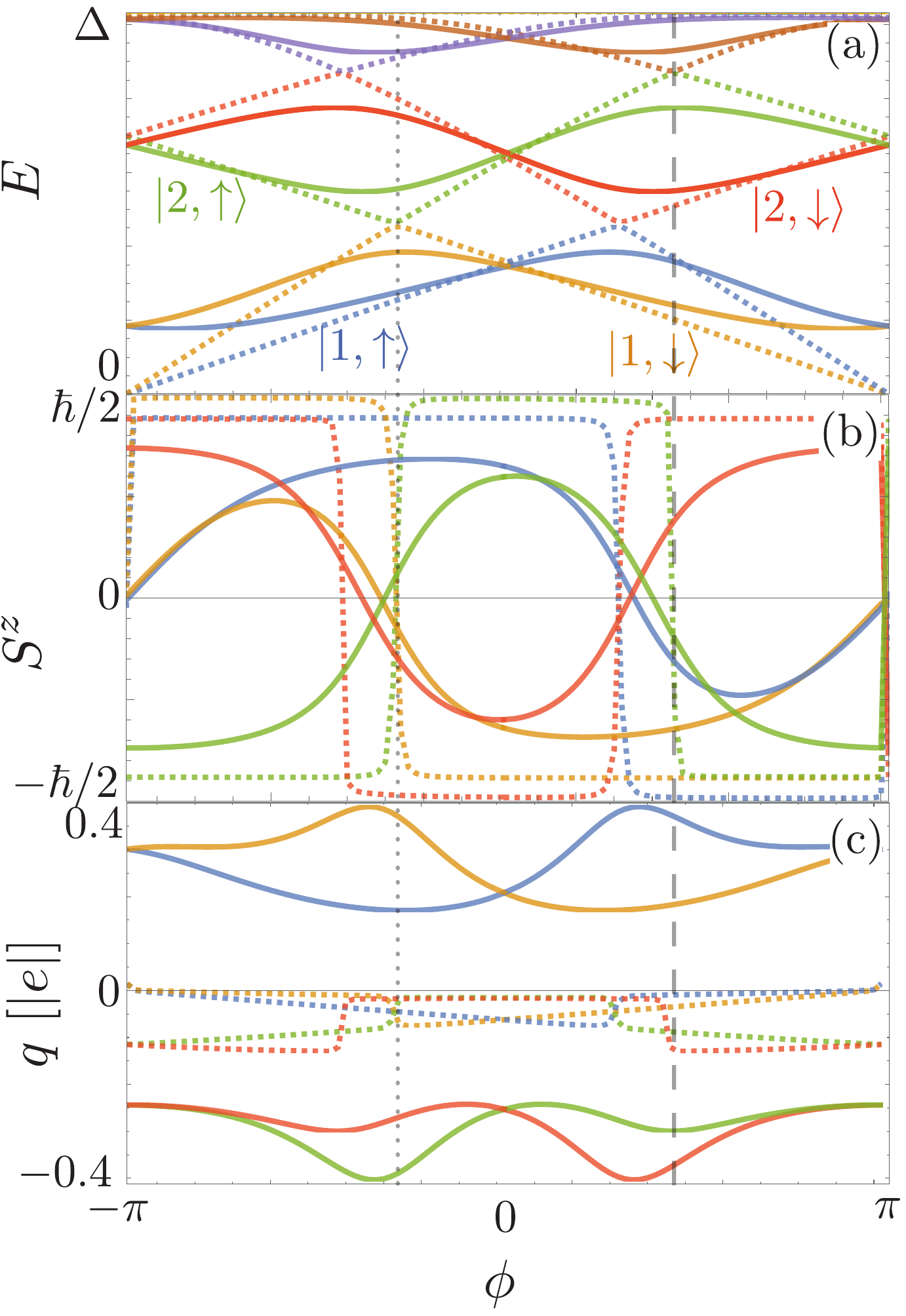}
\caption{A typical Andreev spectrum (a) using $5\times10^2\Delta=10^2\mu'=10^2\alpha'=10(\mu+2t)=10\alpha=t=1$, $L=300$, and $M=1000$  with perfect (dashed), $r=0$, and imperfect (solid), $r=0.03$, transmission between the semiconductor and superconductor; $B^z=0$. Magnetic moment (b), $S_z$, and charge (c), $q$, of the corresponding Andreev states are plotted as a function of $\phi$ where the color and line style coincide with the energies in panel (a). The vertical dotted and vertical dashed lines corresponds to two values of $\phi$ at which Andreev states dispersing with opposite slopes cross.}
\label{ASQ_spec}
\end{figure}

Reintroducing superconductivity to the model, we generate an Andreev spectrum by dividing the chain into three parts in which the superconducting parameters vary by site,
\begin{equation}
(\Delta_j,\phi_j) = 
\left\{
    \begin{array}{lr}
        (\Delta,-\phi/2), & j\leq M\\
        (0,0), & M<j\leq M+L\\
        (\Delta,\phi/2), & j> M+L
    \end{array}
\right.\,,
\end{equation}
and setting $t_j=t[1-(\delta_{j,M}+\delta_{j,M+L})r]$, to reflect a decrease in transmissivity at the superconducting interface with reflectivity $r$ bounded by $0\leq r\leq1$. We plot a typical subgap spectrum as a function of $\phi$ in Fig.~\ref{ASQ_spec}(a), using $5\times10^2\Delta=10^2\mu'=10^2\alpha'=10(\mu+2t)=10\alpha=t=1$, $L=300$, and $M=1000$, with perfect (dashed) and imperfect (solid) transmission between the semiconductor and superconductor by taking $r=0$ and $r=0.03$, respectively. While these parameters are within an order of magnitude of typical experimental parameters, they are chosen to illustrate the salient properties of the Andreev states; we defer a detailed discussion of realistic parameters and how well they predict experimental results until Sec.~\ref{sec:realistic}. 

In Fig.~\ref{ASQ_spec}(a), there are six states within the gap. When $r=0$, the Andreev states linearly disperse with $\phi$ at a slope which is a function of the Fermi velocity.\cite{reynoso2012spin, tosi2019spin} When $r=0.03$, the in-gap states can be grouped into doublets: two in a low-energy doublet, $|1,\uparrow\rangle$ and $|1,\downarrow\rangle$, two in a mid-energy doublet, $|2,\uparrow\rangle$ and $|2,\downarrow\rangle$, and two in a high-energy doublet $|3,\uparrow\rangle$ and $|3,\downarrow\rangle$. Using $\nu$ to index the subgap states, the eigenkets are related to $C_j$ by $\sum_{j=1}^N\psi_{j\nu}\cdot C_j^\dagger|0\rangle$ where $\psi_{j\nu }$ are eight-dimensional spinors in the space of spin, band, and particle-hole, and $\Psi_{\nu}=(\psi_{1\nu},\ldots,\psi_{N\nu})$ is the the $8N$-dimensional eigenstate of Eq.~(\ref{HTB}). While the total number of states in the gap is proportional to the ratio of $L$ to superconducting coherence length, each doublet has two states which are Kramers partners and degenerate at $\phi=0$ or $\phi=\pi$. As one can verify upon comparing the dashed and solid lines of the Andreev spectrum, normal scattering from the superconducting leads results in anticrossings at, for instance, the values of $\phi$ indicated by the vertical dotted and dashed lines in Fig.~\ref{ASQ_spec}. However, even with perfect Andreev reflection, there exists a small but finite anticrossing of Andreev due to the underlying quadratic dispersion [unresolvable in Fig.~\ref{ASQ_spec}(a)].

\subsection{Spin and charge of the Andreev states}

In this work, we will focus on how dephasing occurs via coupling of the environment to the spin and charge of the Andreev levels. 
As such, understanding how the spin and charge character change with external control knobs is a crucial first step. 
In the second-quantized formulation of the tight-binding model, the spin operator at site $j$ is $\textbf{S}_j=C_j^\dagger\textbf{S} C_j$ where $\textbf{S}=(\sigma^x\tau^z,\sigma^y,\sigma^z\tau^z)$ is the spin operator generalized to Nambu space. Likewise, the generalized charge operator at site $j$ is $q_j=C_j^\dagger\tau^z C_j$. 

We plot the spin along the $z$ axis [Fig.~\ref{ASQ_spec}(b)], $S^z_\nu=\sum_j\langle\nu|S^z_j|\nu\rangle$\footnote{According to the form found in Eq.~\ref{HTB}, the eigenstates of have spin only along the $z$ axis.} as a function of phase difference. The phase dependence of the spin can be most easily understood in the case of $r=0$. Upon comparing the spectrum with the spin, we observe that the lowest energy Andreev states with a smaller positive slope (dashed blue curve for $\phi<0$) and larger positive slope (dashed orange curve for $\phi<0$) have spin $S_z\approx\hbar/2$ and $S_z\approx0.9\hbar/2$, respectively.  Moreover, the sign of the spin of the Andreev states is locked to the slope. By comparing this to the Fig.~\ref{infspec}, notice that these spin values coincide with the spin of the bands with larger and smaller Fermi velocity, respectively; it is evident that the spin of the Andreev states is informed by the filling of the underlying semiconducting bands and, consequently, can be controlled via an applied gate voltage. This intuition can be likewise applied to the spin of the Andreev states with finite normal scattering from the superconductor [solid curves in Fig.~\ref{ASQ_spec}(b)]. That is, the spin has the same qualitative dependence on the phase difference but the transitions between positive and negative spin is smoothed by the mixing of the Andreev bands. Evidently, at $\phi=0$, time-reversal symmetry is preserved and each pair of degenerate states must necessarily have opposite spin and, empirically, opposite slope.

In Fig.~\ref{ASQ_spec}(c), we plot the charge of the Andreev states, $q_\nu=\sum_j\langle\nu|Q_j|\nu\rangle$. When $r=0$, the charge is nearly zero. In an infinitely-long junction in which the spectrum is linearly dispersing, perfect Andreev reflection yields Andreev states that are equal parts electron and hole, i.e. zero charge.\cite{nazarovBK09} Because our system approximates these conditions, perfect transparency results in Andreev states with nearly zero charge. In contrast, when $r=0.03$, the Andreev states have a relatively large charge compared to the $r=0$ case. This is a consequence of the otherwise zero-charge states hybridizing into finite-charge states due to the imperfect boundary transparency.

\begin{figure}
\includegraphics[width=\columnwidth]{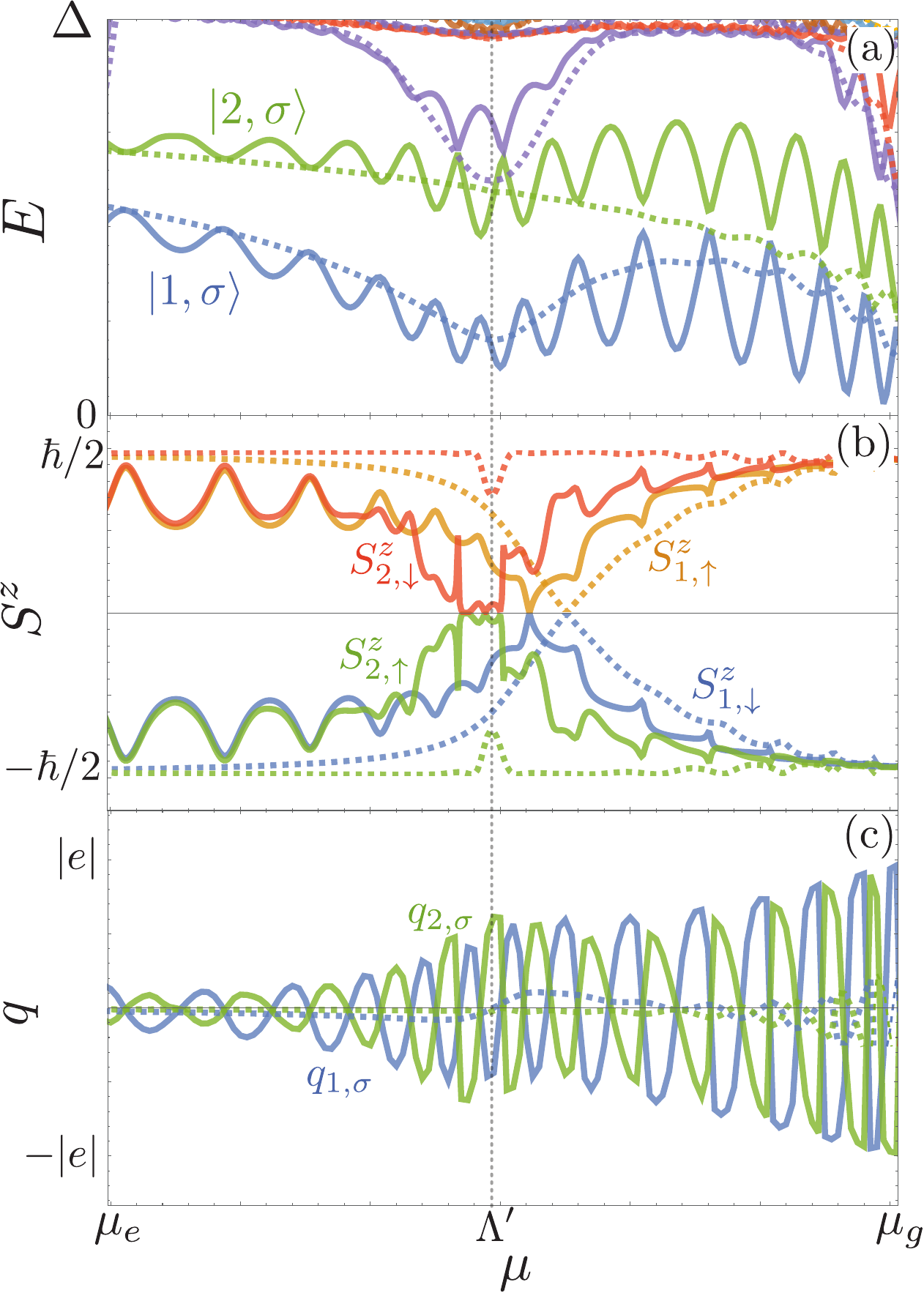}
\caption{Andreev spectrum (a) and associated spin (b) and charge (c) as function of $\mu$, using the same parameters as used in Fig.~\ref{ASQ_spec} and fixing $\phi=0$ with the exception of $\mu$ which is the abscissa. The range of abscissa corresponds roughly to the bottom of the high energy semiconductor subband, $\mu_e=\Delta-\alpha$, and the bottom of the low energy semiconductor subbband, $\mu_g=\delta\Lambda-\Delta-\alpha$. Again, dashed (solid) lines correspond to $r=0$ ($r=0.03$). Because time reversal symmetry is preserved, the energies and electric charges are equal for the $\sigma=\uparrow$ and $\downarrow$ states. The vertical dashed line corresponds to $\mu=\Lambda'$ as in Fig.~\ref{infspec}.}
\label{ASQ_specV}
\end{figure}

Motivated by the observation that the Andreev levels inherit properties from the underlying bands of the semiconductor, we plot the spectrum, spin, and electric charge as a function of the filling, $\mu$, and fixing $\phi=B^z=0$ (Fig.~\ref{ASQ_specV}); the range of $\mu$ is chosen to be roughly from the bottom of the high-energy band, $\mu=\mu_e$, to the bottom of low-energy band, $\mu=\mu_g$. Because the in-gap state energies depend on the Fermi velocity, the spectrum is a function of $\mu$. For $r=0$, we observe a somewhat dramatic dip in the Andreev energies [dashed curves Fig.~\ref{ASQ_specV}(a)] when the Fermi velocity of the semiconducting band is at local minimum (which we define as $\mu=\Lambda'$). Because $\phi=0$, each energy level in the spectrum is doubly degenerate and, consequently, those states have opposite spins [dashed Fig.~\ref{ASQ_specV}(b)]. The magnitude of spin of the lower energy band can vary between $0$ and $\hbar/2$ while the spin of the higher energy band is roughly constant. This is consistent because the spin of the inner branch of the semiconducting states at the Fermi energy [Fig.~\ref{infspec}(b)] varies from $0$ to $\hbar/2$ while the spin of the outer branch is roughly constant over the range of $\mu$ in Fig.~\ref{ASQ_specV}(b). Because the charge is the response of energy to a change in potential, $q$ qualitatively follows $dE/d\mu$: the blue dashed curve is positive (negative) for $\mu>\Lambda'$ ($\mu<\Lambda'$) while the green dashed curve is always negative [Fig.~\ref{ASQ_specV}(c)]. For finite normal reflection ($r=0.03$) the results are qualitatively the same but with additional oscillatory modulation of the spectrum, spin, and charge [Fig.~\ref{ASQ_specV} (solid curves)]. One can show that finite normal reflection mixes Andreev states whose slopes differ in both sign and magnitude, e.g. $|1,\downarrow\rangle$ and $|2,\uparrow\rangle$ \cite{parkPRB17}. This matrix element has a phase which depends on the difference between Fermi momenta. Because the Fermi momentum depends on the filling, the splitting oscillates as a function of $\mu$. Similarly, because the Andreev states are a mixture of two different spins which depends on the same matrix element, the effective spin of the Andreev states also oscillates [Fig.~\ref{ASQ_specV}(b) (solid curves)]. Analogous to the $r=0$ case, the charge in the $r=0.03$ case qualitatively follows $dE/d\mu$ [Fig.~\ref{ASQ_specV}(c) (solid curves)].

\begin{figure}
\includegraphics[width=\columnwidth]{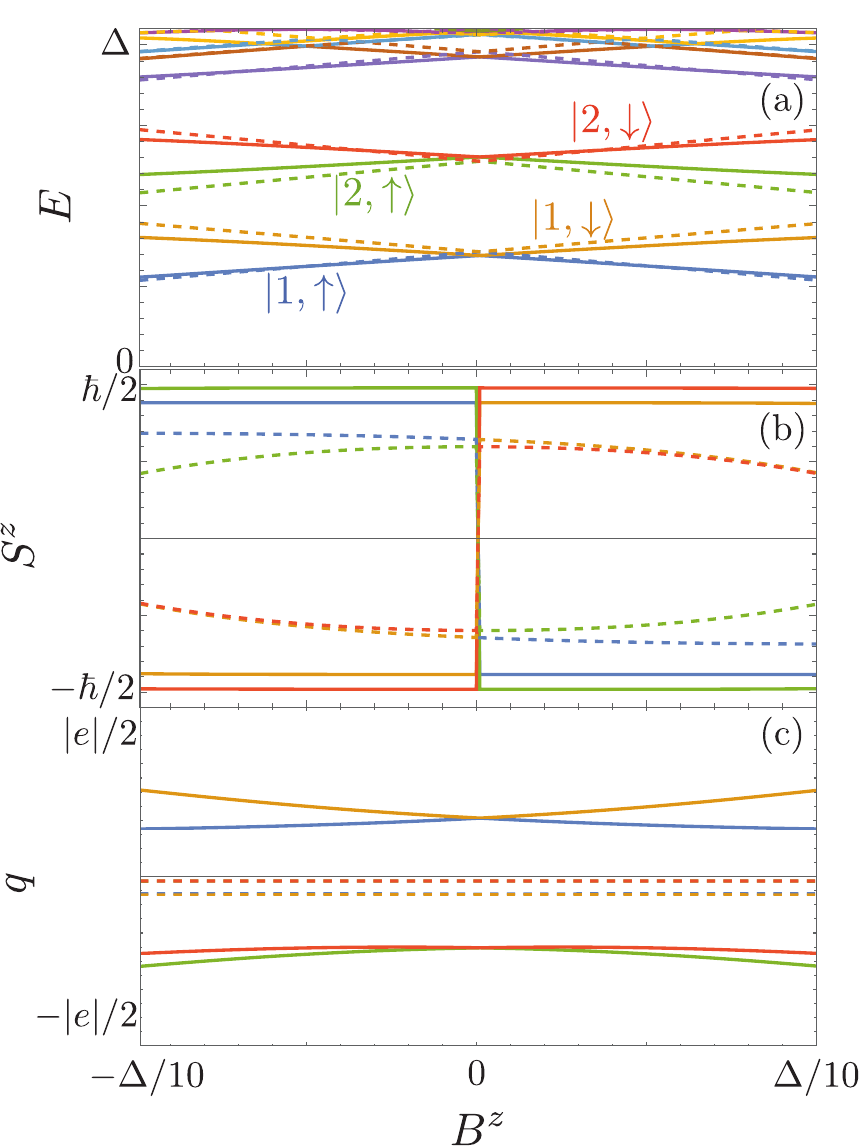}
\caption{Andreev spectrum (a) and associated spin (b) and charge (c) as function of $B^z$, fixing $\phi=0$ and the same parameters as used in Fig.~\ref{ASQ_spec}, with the exception of $B^z$ which is the abscissa. Again, dashed (solid) lines correspond to $r=0$ ($r=0.03$). Because time reversal symmetry is preserved, the energies and electric charges are equal for the $\sigma=\uparrow$ and $\downarrow$ states.}
\label{ASQ_specB}
\end{figure}

Another way to tune the spectrum, both experimentally and within our simulation, is to apply a magnetic field. Upon applying a magnetic field along the polarization axis of the spin-orbit interaction, the spectrum disperses according to the effective $g$-factor [Fig.~\ref{ASQ_specB}(a)]. Notice that, because the states have different spins, the effective $g$-factors differ. While the spin of that states is effectively unchanged [Fig.~\ref{ASQ_specB}(b)], the charge changes according to the energy [Fig.~\ref{ASQ_specB}(c)].

While this section was largely focused on the \textit{total} spin and charge of the individual Andreev states, in our analysis below, an important quantity is the \textit{local} difference in spin and charge between Andreev levels. We define the average local spin and charge difference between states $\nu$ and $\lambda$ as $\mathcal S^z_{\nu\lambda}=\sqrt{\sum_j(\langle\nu|S^z_j|\nu\rangle-\langle\lambda|S^z_j|\lambda\rangle)^2}$ and $\mathcal Q_{\nu\lambda}=\sqrt{\sum_j(\langle\nu|Q_j|\nu\rangle-\langle\lambda|Q_j|\lambda\rangle)^2}$, respectively, where the summation is taken over all sites.

\section{Models of dephasing noise}
\label{sec:Noise Models}

\subsection{Telegraph noise}

Telegraph noise from two level fluctuators is known to contribute to decoherence in both semiconducting spin systems and superconducting qubits\cite{hanson2007spins, krantz2019quantum, Ithier2005decoherence}. Consider $N_t$ two level systems which randomly fluctuate, i.e. exhibit telegraph noise, with some average frequency, $f$ which is sampled from a $1/f$ probability distribution with low-energy (high-energy) cutoff at $\Lambda_\textrm{IR}$ ($\Lambda_\textrm{UV}$). Assuming the fluctuators act locally, their microscopic coupling to the electrons is
\begin{equation}
H_{tb}=\sum_{n=1}^{N_t}\xi_n(t) F_n D_{j_n}\,
\end{equation}
where $\xi_n(t)$, reflecting the time-dependent state of the $n$th two-level system, takes values $\pm1$ and $F_n$ is a phenomenological parameter which captures the coupling between the $n$th fluctuator at site $j_n$ and local observable which has the general form $D_{j_n}=\sigma^\beta\rho^\gamma\tau^\delta$ where $\beta,\gamma,\delta=0,x,y,z$. While we could consider any local observable, we are most interested when the observables are charge and spin along the $z$ axis wherein $D_{j}=C^\dagger_j\tau^z C_j$ and $D_{j}=C^\dagger_j\sigma^z\tau^z C_j$, respectively. Projecting this Hamiltonian onto $N_A$ in-gap states we obtain 
\begin{equation}
H_{tb}=\sum_{n=1}^{N_t}\xi_n(t)F_n \boldsymbol{\mathcal{D}}_n\,,
\label{Htqb}
\end{equation}
where $(\boldsymbol{\mathcal{D}}_n)_{\nu\lambda}=\langle\nu|D_{j_n}|\lambda\rangle$. Taking $\nu$ and $\lambda$ to be the lowest energy in-gap states and projecting $H_A$ onto them, we obtain the Hamiltonian
\begin{equation}
H_{tq}=\omega_q\eta_z+F(t)\eta_z\,,
\label{Htq}
\end{equation}
which is reminiscent of a semiconducting spin qubit in a random magnetic field, $F(t)$. Notably, because of spin-orbit interaction, the fluctuations in the qubit frequency can be caused by either electric (charge-coupled) or magnetic (spin-coupled) noise. 

\subsection{Nuclear Spin bath}

One source of decoherence which is present in quantum dots are nuclear spins. Microscopically, the interaction between electrons and nuclei is governed by the hyperfine Fermi contact interaction. We encode the coupling of~$N_I$ nuclei to the electrons in our tight-binding model using the interaction
\begin{equation}
    H_{Ab}=\sum_{n=1}^{N_I} A_n \textbf{S}_{j_n}\cdot \textbf{I}_n\,,
    \label{Hhf}
\end{equation}
where~$N_I$,$\boldsymbol{I}_n$, and $j_n$ are the number of nuclei, their magnetic moments, and position of the $n$th nucleus, respectively. $A_n$ is a phenomenological parameter characterizing coupling of the $n$th nucleus to the electron at site $j$. When the spacing between the continuum states and Andreev states is much larger than $A_n$, we use first order perturbation theory to project Eq.~(\ref{Hhf}) onto the $N_A$ subgap states,
\begin{equation}
    H_{qb}=\sum_{n=1}^{N_A} A_n \boldsymbol{\mathcal{S}}_n\cdot \textbf{I}_n\,.
    \label{Hqb}
\end{equation}
$(\boldsymbol{\mathcal{S}}_{n})_{\nu\lambda}=\langle\nu|\hat{\textbf{S}}_{j_n}|\lambda\rangle$ is an $N_A\times N_A$-matrix describing both a shift in energies of the Andreev states and transitions between those states as a result of the nuclei. For $N_A=2$, Eq.~(\ref{Hqb}) looks suggestive of an electron confined to a quantum dot interacting with nuclei via the hyperfine interaction wherein it reduces to $H_{qb}=\sum_{n}\boldsymbol{\eta}\cdot\hat{A}_n\cdot \textbf{I}_{n}$, with $\boldsymbol{\eta}=(\eta_x,\eta_y,\eta_z)$ the Pauli matrices acting on the two-dimensional subspace of states. To simplify our system and get intuition for the form of $\hat{A}_n$, we restrict ourselves to the lowest two Andreev states.\footnote{Overhauser fields perpendicular to the direction of spin polarization can induce transitions between Andreev states when the nuclear Larmour frequency is on or nearly resonant with Andreev level spacing. Moreover, parallel mode Overhauser fields can induce transitions between Andreev states of the same spin. In order to focus on decoherence properties of Andreev qubits, we do not consider qubits composed of states with the same spin, e.g. the lowest and fourth lowest Andreev state, hence the restriction to the three lowest energy Andreev states.} In this case, $\hat{A}_n=A_n\textrm{diag}[\textrm{Re}(\langle1,\uparrow|\hat{S}^x_{j_n}+\hat{S}^y_{j_n}|1,\downarrow\rangle),\textrm{Im}(\langle1,\uparrow|\hat{S}^x_{j_n}+\hat{S}^y_{j_n}|1,\downarrow\rangle),\langle1,\uparrow|\hat{S}^z_{j_n}|1,\uparrow\rangle-\langle1,\downarrow|\hat{S}^z_{j_n}|1,\downarrow\rangle]$ is a diagonal matrix which looks similar to the hyperfine interaction of an electron confined to a quantum dot but dressed according to the Andreev state wavefunctions. Because we have chosen, without loss of generality, the spin orbit polarization axis to be along $z$, transitions between states are generated by nuclear magnetic moments in the $xy$ plane while nuclear magnetic moments along the $z$ axis shift the Andreev state energies. Analogously to the case of semiconducting spin qubits, when the nuclear Larmour frequency is far off resonant from the Andreev spin qubit, relaxation is suppressed and the effective hyperfine interaction reduces to 
\begin{equation}
    H_{qb}=\sum_{n=1}^{N_I} (\langle\nu|\hat{S}^z_{j_n}|\nu\rangle-\langle\lambda|\hat{S}^z_{j_n}|\lambda\rangle) \eta^z I_n^z\,.
    \label{Hqbred}
\end{equation} 
Because decoherence times are typically much shorter than relaxation times in experiments,\cite{hays2021CoherentASQ,pita2023direct} we focus on the limit where Eq.~(\ref{Hqbred}) is a good approximation for the qubit-bath interaction. 

Additionally, in order to generate decoherence, we assume the bath undergoes internal dynamics according to 
\begin{align}
    H_{b0}&=\sum_n\omega_n \textbf I_{nz}\,,\nonumber\\
    H_{bb}&=\sum_{mn} b_{mn}  \textbf I_{n} \cdot \textbf I_{m}\,,
\end{align}
where $\omega_n$ is the characteristic frequency of the $n$th nuclei and $b_{mn}$ describes the dipole interaction between nuclei $m$ and $n$. While the nuclei can in general be a large spin, e.g. the magnetic moment of In is 7/2, we restrict $|\textbf{I}_n|=1/2$ to be a phenomenological two-level system wherein $\omega_n$ captures any splitting between the two levels arising from, for example, strain, applied magnetic field, etc.

Lastly, projecting Eq.~(\ref{HTB}) onto the subgap states we obtain $(H_A)_{\nu\lambda}=E_\nu\delta_{\nu\lambda}$ where $E_\nu$ is the energy of the $\nu$th Andreev state. Using only two of those states and defining their difference in energy to be $\omega_q=(E_\nu-E_\lambda)/2$, we obtain the full, effective qubit-bath Hamiltonian,
\begin{equation}
    H_\textrm{eff}=\omega_q\eta_z+\sum_n\boldsymbol{\eta}\cdot\hat{A}_n\cdot \textbf{I}_{n}+\sum_n\omega_n  I_{nz}+\sum_{mn} b_{mn}  \textbf I_{n} \cdot \textbf I_{m}\,.
    \label{Heff}
\end{equation}
In general, $\omega_q$ and $\hat A_n$ depend on the phase difference, applied magnetic field, and qubit encoding, i.e. which two subgap states were chosen, so the decoherence time should depend on these external parameters and qubit choices. 

\section{Understanding Decoherence}
\label{sec:Decoherence}

To simulate decoherence, we initialize the qubit into an equal superposition of the logical Andreev states and the bath into a random state. We evolve the initial states for time $\tau/2$, perform a $\pi$ pulse on the qubit, and evolve for another $\tau/2$ (imitating a Hahn echo decoherence experiment). The off diagonal elements of the qubit density matrix are extracted and plotted as a function of $\tau$. Upon fitting this curve to $\epsilon\exp[-(\tau/T_{2E})^\kappa]+\lambda$, we can extract the echo coherence time, $T_{2E}$. The simulation is identical to extract the Ramsey coherence time, $T_{2R}$, absent the $\pi$-pulse.

\subsection{Decoherence due to telegraph noise}

Using the model parameters of Sec.~\ref{sec:Andreev states}, we first calculate the decoherence of Andreev levels coupled to a bath of TLSs. We are interested in two cases: when the fluctuators couple to the local (1) electric and (2) magnetic noise, $D_{j_n}=C_j^\dagger\tau^z C_j$ and $D_{j_n}=C_j^\dagger\sigma^z\tau^z C_j$, respectively. In both cases, we consider $N_t=100$, $F_n=F$ and $\Lambda_\textrm{UV}=10^2\Lambda_\textrm{IR}$. To accurately sample the rate of fluctuating noise, we choose the time step in which the system is evolved such that it is much smaller than the high frequency cutoff, $\Lambda_\textrm{UV}^{-1}/10$. Moreover, demanding the coherence of the qubit encoded in the lowest two states to be of order 1~ns fixes the product of $F$ and $\mathcal D_n$; because $\mathcal D_n$ is different for electric and magnetic noise, in general $F$ will be different as well. Simultaneously, this fixes $\Lambda_\textrm{UV}=1$~GHz and $\Lambda_\textrm{IR}=10$~MHz. 

\begin{figure}
\includegraphics[width=\columnwidth]{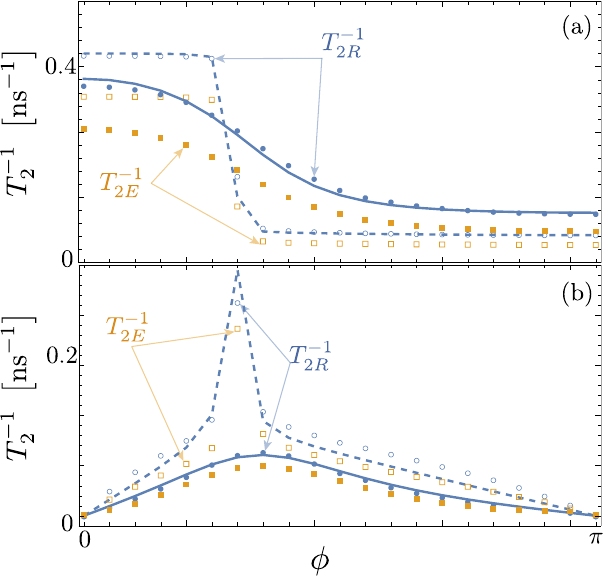}
\caption{Dephasing rate as a function of phase difference, $\phi$, using the same parameters as Fig.~\ref{ASQ_spec}, due to magnetic (a) and electric fluctuators (b); the open markers use $r=0$ and the closed markers use $r=0.03$. The Ramsey dephasing rate (echo dephasing rate), $1/T_{2R}$ ($1/T_{2E}$), are marked with blue circles (orange squares). The local difference in magnetic moment (a) and charge (b) is denoted by a dashed curve for $r=0$ and solid curve for $r=0.03$.}
\label{T2a}
\end{figure}

We plot the dephasing rate, i.e. the inverse of the coherence time in the absence of energy relaxation, as a function of phase difference when the qubit is coupled to magnetic noise [Fig.~\ref{T2a}(a)] where we have used $F=200$~GHz.\footnote{While this number appears large, we note that, because the fluctuator couples to the qubit only at a single site, the effective coupling between the fluctuator and qubit is $\sim100$~MHz. Moreover, to fix the coherence time while increasing the number of fluctuators, the product of $B$ and $sqrt{N_{t}}$ should remain constant. Consequently, in a realistic system where $N_t\sim10^5$, we expect the effectively coupling between fluctuators and the qubit to be $\sim10$~MHz which is experimentally reasonable.} When $r=0$, the coherence time precipitously drops as the phase difference goes through $\phi\approx\pi/4$, corresponding to the crossing of the second and third lowest energy levels [Fig.~\ref{ASQ_spec}(a)]. This is a consequence of the difference in spin of the positively and negatively dispersing Andreev states [Fig.~\ref{ASQ_spec}(b)]. Quantitatively, we find that the average local difference in spin [Fig.~\ref{T2a}(a)], $\mathcal S^z_{(1,\uparrow),(1,\downarrow)}$,  is proportional to $1/T_{2R}$. $T_{2E}$ follows the same qualitative shape as $T_{2R}$, though the magnitude is roughly twice as large. For $r=0.97$, the Ramsey coherence time gradually increases as $\phi$ is increased from $0$ to $\pi$ as a result of the gradual change in spin [Fig.~\ref{ASQ_spec}(b)] which is a result of the anticrossing in the spectrum [Fig.~\ref{ASQ_spec}(a)].

When the qubit is coupled to charge noise [Fig.~\ref{T2a}(b)] and choosing $F=4$~THz, $1/T_{2R}$ is zero when time reversal symmetry is preserved ($\phi=0,\pi$) because the spatial profile of the charge is identical between the two lowest energy states. As phase increases, the difference in charge increases which matches the decrease in $T_{2R}$. At the point where the states cross in the spectrum ($\phi\approx\pi/4$), there is a decrease in coherence time. Past that point, the coherence time increases linearly in accordance with the charge difference of the states. Analogous to magnetic noise, we find that the average local difference in charge, $\mathcal Q_{(1,\uparrow,1,\downarrow)}$, matches the inverse coherence time well. Likewise, $T_{2E}$ follows similar qualitative behavior as $T_{2R}$. When the transparency is imperfect ($r\neq 0$), $T_2$ is a continuous function of $\phi$. Similar to the case of perfect transparency, $1/T_{2R}\sim1/T_{2E}$ which is proportional to the average local difference in charge. 

\begin{figure}
\includegraphics[width=\columnwidth]{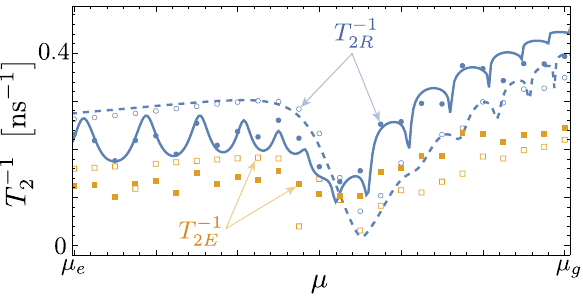}
\caption{Dephasing rate due to magnetic fluctuators as a function of filling, $\mu$, using the same parameters as Fig.~\ref{ASQ_spec} at $\phi=0$; the open markers use $r=0$ and the closed markers use $r=0.03$. The simulated Ramsey dephasing rate (echo dephasing rate), $1/T_{2R}$ ($1/T_{2E}$), are marked with blue circles (orange squares). The local difference in magnetic moment denoted by a dashed curve for $r=0$ and solid curve for $r=0.03$. The lower and upper bound of $\mu$ are the same as in Fig.~\ref{ASQ_specV}.}
\label{T2V}
\end{figure}

In Fig.~\ref{T2V}, we plot the decoherence times as a function of the filling, $\mu$, when coupled to magnetic noise. Strikingly, when $r=0$, there exists a filling for which the dephasing rate is minimized. For this value of $\mu$, the Fermi energy is near the anticrossing of the semiconducting bands wherein the total spin in the semiconducting sector is zero (Fig.~\ref{infspec}). Although the local spin of the Andreev states remains finite, it is diminished and hence decreases the dephasing rate. Although slightly obscured by oscillations, there exists an analogous decrease in dephasing rate for $r=0.03$. Similar to the phase-dependent coherence times, the filling-dependent $1/T_2$ is proportional to the average local difference in spin. The inverse of the echo times, $1/T_{2E}$, qualitatively follows $1/T_{2R}$ with a smaller magnitude. While there is some discrepancy in the $\mu$-dependent behavior between the decoherence times, we can partially ascribe this to some poor functional fitting of the coherence curves in the time-domain. Because time reversal symmetry is preserved, the lowest energy states are Kramer's partners and consequently have identical total charge [Fig.~\ref{ASQ_specV}(c)] and charge density. As a result, the coherence time diverges (figure omitted) for all values of $\mu$ when the Andreev states are coupled to charge noise.

\begin{figure}
\includegraphics[width=\columnwidth]{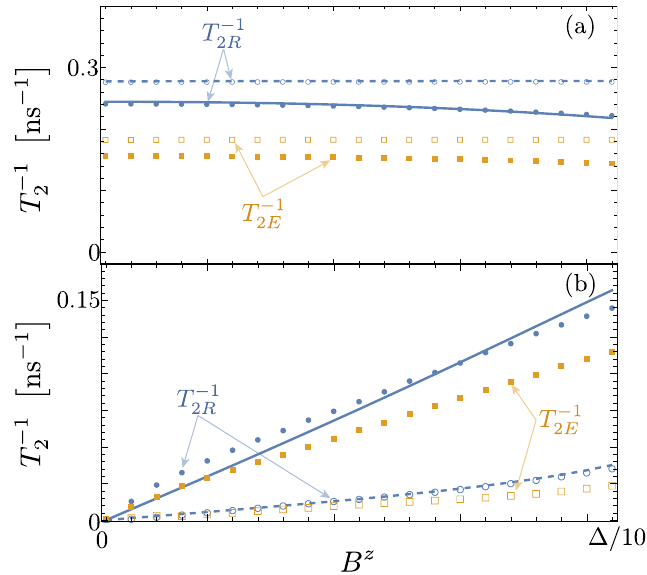}
\caption{Dephasing rate as a function of magnetic field, $B^z$, using the same parameters as Fig.~\ref{ASQ_spec}, due to magnetic (a) and electric fluctuators (b); the open markers use $r=0$ and the closed markers use $r=0.03$. The simulated Ramsey dephasing rate (echo dephasing rate), $1/T_{2R}$ ($1/T_{2E}$), are marked with blue circles (orange squares). The local difference in magnetic moment (a) and charge (b) denoted by a dashed curve for $r=0$ and solid curve for $r=0.03$.}
\label{T2B}
\end{figure}

The dephasing rate as a function of $B^z$ is roughly constant when the qubit is coupled to magnetic noise [Fig.~\ref{T2B}(a)]. Again, the average local difference in spin is proportional to $1/T_2$. Contrasting with the dephasing as a result of electric noise [Fig.~\ref{T2B}(b)], $1/T_2$ strongly depends on the magnetic field. Specifically, the coherence time diverges as $B^z$ approaches $B^z=0$, i.e. the system becomes time-reversal symmetric. 

To conclude this section we summarize our results: By changing the externally controllable parameters, $\phi$, $\mu$, and $B^z$, one can strongly enhance or dimish the coherence times. Moreover, the underlying local electric charge and spin of the Andreev states appears to be, at the very least, a good qualitative indicator of the dependence of the coherence time on system parameters. As we see in the next section, this informs our intuition about the decoherence mechanisms in realistic systems.

\section{Realistic model, comparing to experiment}

\label{sec:realistic}

While the system parameters used in the previous section were instructive in understanding the behavior of parameter-dependent decoherence and its relation to the underlying properties of the Andreev bound states, the Hamiltonian values used in Eq.~(\ref{HTB}) do not result in a particularly close match to experimentally measured spectra~\cite{tosi2019spin, hays2021CoherentASQ, pita2023direct}. To remedy this, we now tune our Hamiltonian parameters to obtain a rough match with the spectra measured in Ref. \onlinecite{pita2023direct}. 

\begin{figure}
\includegraphics[width=\columnwidth]{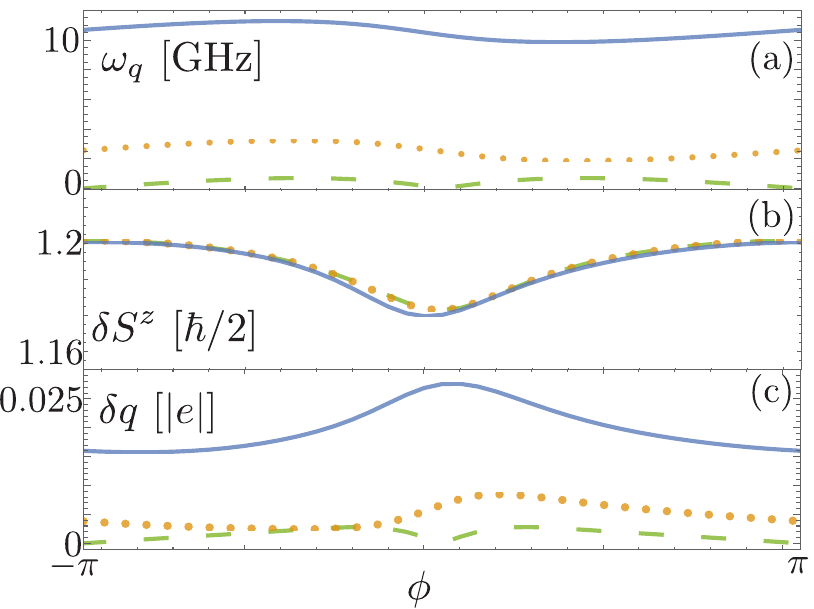}
\caption{Realistic qubit frqency (a), $\omega_q$, difference in total spin (b), $\delta S^z$, and difference in total charge (c), $\delta q$, using $\Delta/t=10^{-3}$, $\mu/t=-2.01$,  $\alpha/t\approx0.058$, $\mu'/t\approx0.001$, $\alpha'/t\approx0.008$, $r=0.05$, $L=69$, and $M=500$, corresponding to a semiconducting region of length $200$~nm, width of $100$~nm and spin-orbit length of $50$~nm. The solid blue, dotted orange, and dashed green correspond to three different values of magnetic fields $B^z=70$~mT, $B^z=17$~mT, and $B^z=0$~mT, respectively.}
\label{spec}
\end{figure}

To begin, we take the length of the wire to be $200$~nm and the width to be $100$~nm, the latter of which determines $\mu'$.\cite{parkPRB17} Using a spin-orbit length of $50$~nm, which is similar to experiment\cite{roulleauPRB10,hernandezPRB10}, $\alpha$ and $\alpha'$ are determined. Next, we choose the gap to be $10^{-3} t$ and, using a gap size of 50~GHz, this sets the units in Eq.~(\ref{HTB}). To fix $\mu$ and $r$, we match the qubit frequency as a function of $\phi$ to Ref.~\onlinecite{pita2023direct}; that is, the value of $\mu$ will inform the Fermi velocities, which in turn gives the change in the qubit frequency as $\phi$ is varied. When $B^z/t\approx2\times10^{-4}$ and $\phi=0$, the qubit frequency is slightly largely than 10~GHz. Upon comparison with Ref.~\onlinecite{pita2023direct}, we identify this value of $B^z$ in our model with an applied magnetic field of 70~mT which resolves a $g$-factor of approximately 10 which fits with experiment. Moreover, decreasing the magnetic field to $17$~mT, we obtain a maximal qubit frequency of $\approx3.3$~GHz at $\phi\approx-\pi/2$ and a minimal qubit frequency of $\approx1.9$~GHz at $\phi\approx\pi/2$ which also matches experiment quite nicely.\footnote{The position of the maximum and minimum qubit frequency can be reversed by effectively reversing the direction of the magnetic field, i.e. $B^z\rightarrow-B^z$.} While these are likely not the only parameters that recover similar spectral properties to experiment, this is one example where they do [Fig. \ref{spec}(a)].

As with the previous section, we are interested in the coherence times as a function of externally-tunable parameters when the qubit is coupled to electric or magnetic noise. Experimentally, the coherence time has been measured as a function of external magnetic field and external flux, corresponding to the parameters $B^z$ and $\phi$ in our model \cite{pita2023direct}. In the experiment, as $B^z$ was varied, the phase difference was fixed so that the qubit frequency was maximized, i.e. $\phi\approx-\pi/2$. As $\phi$ was changed in experiment, the qubit frequency was maintained at approximately $11$~GHz which demands the magnetic field be varied between 65~mT and 70~mT accordingly. When studying the $\phi$-dependent decoherence, we fix $B^z=0$~mT or $70$~mT and, when studying the $B^z$-dependent decoherence, we fix $\phi=0$ or $-\pi/2$. While the values of $B^z=70$~mT and $\phi=-\pi/2$ are consistent with experiment, $B^z=0$ and $\phi=0$ are simpler cases which facilitate better physical intuition. As the experimental dependence of $T_2$ on $B^z$ and $\phi$ is quite flat\cite{pita2023direct}, we expect the fluctuators that result in flatter $T_2$ curves to be a better match with experiment. 

\begin{figure}
\includegraphics[width=\columnwidth]{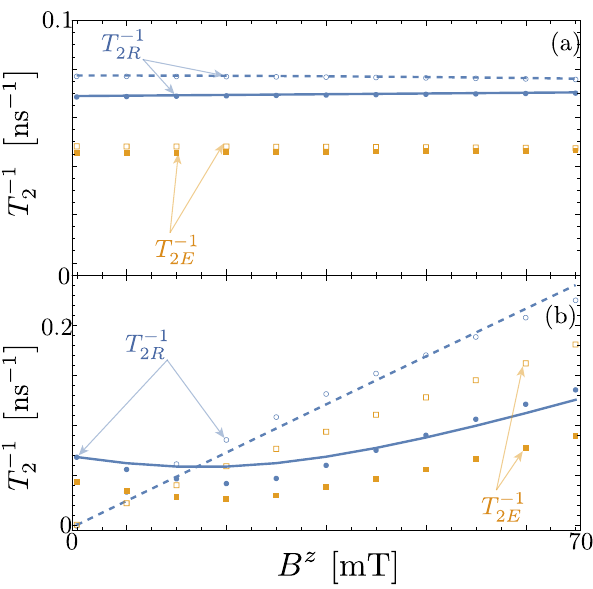}
\caption{Realistic qubit dephasing as a function of $B^z$ for $\phi=0$ (open markers) and $\phi=-\pi/2$ (filled markers) as a result of coupling to magnetic (a) and electric (b) fluctuators using the same parameters as in Fig.~\ref{spec}.}
\label{realTB}
\end{figure}

Consider first the dephasing rate dependence on the magnetic field. In Fig.~\ref{realTB}(a), we plot $1/T_{2R}$ (dashed) and $1/T_{2E}$ (solid) as a function of $B^z$ when the Andreev states are coupled to magnetic noise and for $\phi=0$. The dependence is quite weak and, as with the previous set of parameters, follows the difference in local magnetic moment quite well. For $\phi=-\pi/2$, the dephasing displays similar behavior. When the Andreev states are coupled to charge noise, the dephasing strongly depends on the magnetic field. In particular, when time reversal symmetry is preserved, i.e. $B^z=\phi=0$, the coherence time diverges. When $\phi=0$, the dephasing appears to increase monotonically and roughly linearly with $B^z$. In contrast, for $\phi=-\pi/2$, the dephasing is a nonmonotonic function of $B^z$. 

\begin{figure}
\includegraphics[width=\columnwidth]{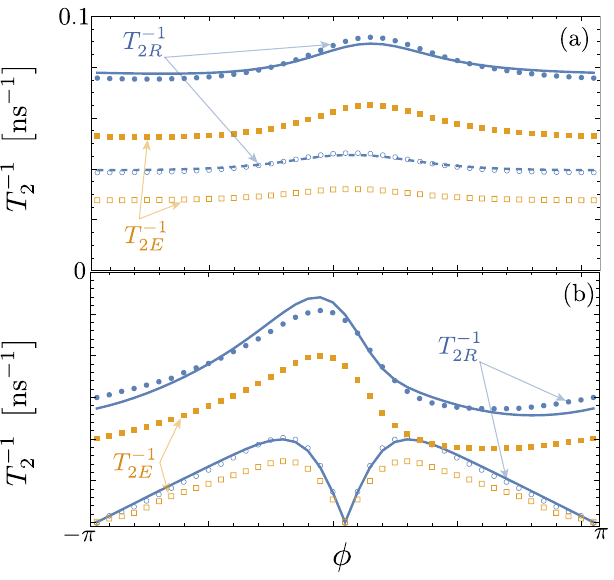}
\caption{Realistic qubit dephasing as a function of $\phi$ for $B^z=0$ (open markers) and $B^z=70$~mT (filled markers) as a result of coupling to magnetic (a) and electric (b) fluctuators using the same parameters as in Fig.~\ref{spec}.}
\label{realTF}
\end{figure}

In Fig.~\ref{realTF}, we plot the dephasing as a function of phase difference. When the Andreev states are coupled to magnetic noise [Fig.~\ref{realTF}(a)] and $B^z=0$ (open markers), both dephasing rates are peaked at $\phi=0$ which correlates well with the difference in local magnetic moment. The analogous dephasing when $B^z=70$~mT, is similar to the $B^z=0$ case but is peaked slightly away from $\phi=0$ which is similarly consistent with the difference in local magnetic moment. When coupled to electric noise [Fig.~\ref{realTF}(b)], the $\phi$-dependence on the dephasing is highly anisotropic and goes to zero, i.e. the coherence time diverges, when time-reversal symmetry is preserved, i.e. $\phi=0,\pm\pi$. As the magnetic field is turned on, the divergence in the coherence time is lifted and the dephasing rate is peaked near $\phi=0$. The local charge difference closely matches the dephasing in both the $B^z=0$ (dashed blue curve) and $B^z=70$~mT (solid blue curve) case.

\begin{table*}[t]
\begin{tabular}{ |P{1.5cm}||P{2cm}||P{2.5cm}||P{2.5cm}||P{2.5cm}|}
 \hline
 \multicolumn{5}{|c|}{$\chi^2$} \\
 \hline
 & electric TLS &magnetic TLS&(short range) nuclear bath &(long range) nuclear bath \\
 \hline
{$T_{2R}(B^z)$}   & 13    &1&1 &1\\
 {$T_{2E}(B^z)$} &   29  & 3 &3 &3 \\
 {$T_{2R}(\phi)$} &8 & 1&1&0\\
 {$T_{2E}(\phi)$}    &35 & 12&7&5\\
 \hline
Total    &85 & 17&12&9\\
 \hline
\end{tabular}
\caption{$\chi^2$ comparison of experimental data extracted from Ref.~\onlinecite{pita2023direct} to our numerically calculated dephasing rates as a result of coupling to electric TLSs, magnetic TLSs, nuclei coupled at individual sites, and nuclei coupled over many sites. For brevity, we have rounded to the nearest whole number.}
\label{chi2}
\end{table*}

How do these different mechanisms compare to experiment? To answer this, we compare our simulations to the $B^z$- and $\phi$-dependent $T_2$ times reported in Ref.~\onlinecite{pita2023direct}. Note that our $T_2$ simulations can be scaled by a global factor by scaling the interactions between qubit and the bath, $F$, and the bath fluctuation rate cutoffs, $\Lambda_\textrm{UV}$ and $\Lambda_\textrm{IR}$. Accordingly, we scale our $B^z$- and $\phi$-dependent Ramsay and echo coherence times to best fit the experimental results according to the $\chi^2$ goodness of fit. The simulated noise source which minimizes the $\chi^2$ is, by definition, a better fit to experiment [Tab.~\ref{chi2}]. We observe, rather generally, that the magnetic noise fits significantly better than the electric noise.

Because we are more confident that magnetic noise is the dominant source of dephasing, we study the effects of a more sophisticated magnetic noise model: a nuclear spin bath. For $A_n$, the coupling between Andreev states and nuclei ($b_{mn}$, the internuclear coupling), we take an average value of 4~GHz (4~GHz) which are normally distributed with a variance of 4~GHz (2~GHz). The variance in $A_n$ phenomenologically captures the positioning of nuclei throughout different points in the crosssection of the wire. Similarly, the variance in $b_{mn}$, phenomenologically captures variations in distances between nuclei. In our simulations, we use $N_I=7$ nuclei whose positions are chosen at random at sites in the semiconducting portion of the wire. We take twenty of these random collections of sites, calculate the dephasing, i.e. the off-diagonal elements of the qubit density matrix, and average the results. Because the difference in local magnetic moments depends on position, we expect this to give a reasonable sample of effective hyperfine interactions.

\begin{figure*}[t]
\includegraphics[width=1.9\columnwidth]{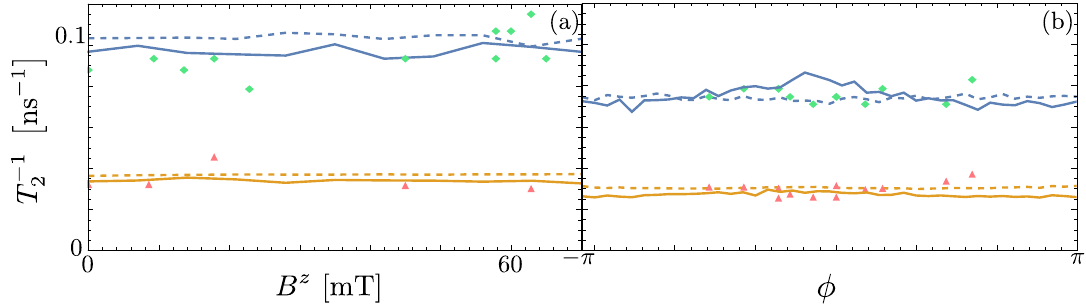}
\caption{Numerically simulated decoherence rates resulting from short range (solid curve) and long range (dashed curve) coupled nuclei as a function of $B^z$ (a) and $\phi$ (b), using the same parameters as in Fig.~\ref{spec}. For comparison, we have overlayed $T_{2R}$ (green diamonds) and $T_{2E}$ (red triangles) experimental data from Ref.~\onlinecite{pita2023direct}.}
\label{T2Rc}
\end{figure*}

In Fig.~\ref{T2Rc} (solid curve), we plot the $B^z$- and $\phi$-dependence of the dephasing and superimpose the experimental data on top of it. Indeed, we find that $1/T_{2R}$ and $1/T_{2E}$ match the experimental results qualitatively and quantitatively well. Similar to the telegraph noise, we can scale the data by scaling the Hamiltonian of the nuclear bath and calculate the best resulting $\chi^2$ [Tab.~\ref{chi2}]. Indeed, this appears comparably as good as the magnetic TLS dephasing model. Nonetheless, notice that the dephasing found in experiment is constant or even dips near $\phi=0$ while the dephasing as a result of magnetic noise, either the fluctuators or the nuclear spin bath, is peaked near $\phi=0$. This is due to the fact that fluctuators and nuclei act locally and so sample the local difference spin of the Andreev states. In contrast, according to Fig.~\ref{spec}(b), the total spin difference of the Andreev states has a dip near $\phi=0$. If a magnetic noise source couples to the Andreev states over a long distance, we expect to see a peak dip at in the corresponding dephasing at $\phi=0$. Plotting the dephasing as a result of nuclei that act over the size of the semiconducting region [Fig.~\ref{T2Rc} (dashed curve)], we find that the dephasing as a function of $\phi$ is largely flat because the change in total spin with phase is quite small but appears to fit the data better qualitatively. Quantitatively, the goodness of fit is slightly better than the local spin fluctuations [Tab.~\ref{chi2}]. Although we have only displayed our results for the nuclear spin bath, we obtain similar parameter dependent coherence times when magnetic TLSs couple to the Andreev states over long distance.

\begin{figure}
\includegraphics[width=\columnwidth]{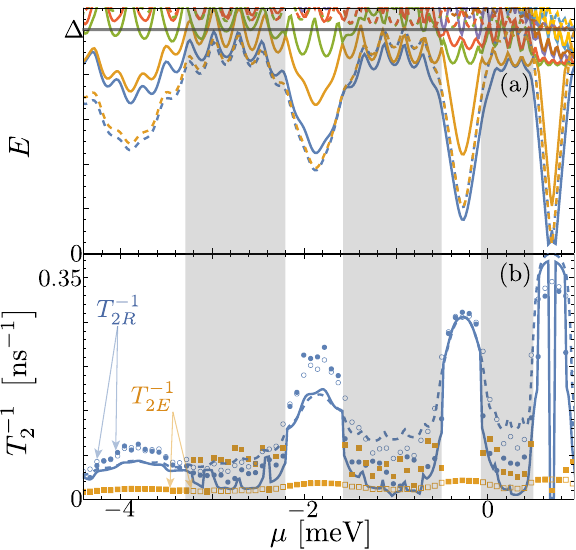}
\caption{Spectrum as a function of $\mu$ where the dashed lines are for $B^z=0$ and $\phi=-\pi/2$ and the solid lines for $B^z=70$~mT and $\phi=0$ but otherwise using the same parameters as Fig.~\ref{spec} (a). Dephasing due to nuclei as a function of $\mu$ for $B^z=0$ and $\phi=-\pi/2$ (open markers) and $B^z=70$~mT and $\phi=0$ (filled markers) (b). Dashed (solid) blue curves in panel (b) are the average local magnetic moment for $B^z=0$ and $\phi=-\pi/2$ ($B^z=70$~mT and $\phi=0$). Grayed areas mark values of $\mu$ in which the in-gap states merge with the continuum.}
\label{specVR1b}
\end{figure}

As the nuclear spins appear to be a possible contributor to decoherence, we consider enhancing (short-range)-nuclear-spin limited $T_2$: As illustrated in the Sec.~\ref{sec:Andreev states}, we found that by changing the filling, $\mu$, one can change the effective spin of the Andreev states. Attempting the same strategy for the realistic model, we first plot the Andreev spectrum as a function of $\mu$ over a filling range of the order 1~meV, for the case of (1) $B^z=0$ and $\phi=-\pi/2$ and (2) $B^z=70$~mT and $\phi=0$. There are a handful of minima in the spectrum which are the result of an anticrossing of particles and holes and have the same origin as the oscillations in Fig.~\ref{ASQ_specV}(b). However, because the length is significantly smaller in this case as compared to the model parameters in Sec.~\ref{sec:Andreev states}, the range in abscissa between local minima is comparatively larger. According to the spectrum, the qubit frequency for $B^z=70$~mT is larger than the qubit frequency for $\phi=-\pi/2$ but the qualitative dependence of the qubit frequency on $\mu$ is similar in both cases. Moreover, there are several values of $\mu$ in which the Andreev states merge with the gap (grey shaded regions in Fig.~\ref{specVR1b}). While we can formally calculate the dephasing rate using the lowest two energy states as a qubit, we have not carefully checked that they are localized states. Moreover, these states that surf the gap edge would be difficult to spectrally access in experiment and focus on the white areas of Fig.~\ref{specVR1b}.

In Fig.~\ref{specVR1b}(b), we plot the dephasing as a result of coupling to the nuclear spin bath when $\phi=-\pi/2$ and $B^z=70$~mT. Observe, for both sets of parameters, that the deeper the state in the gap, the shorter the $T_{2R}$ and $T_{2E}$. Upon plotting the average local magnetic moment difference, we observe that the lower energy states indeed have a larger magnetic moment and, therefore, a higher dephasing rate. Consequently, the dephasing rate can be significantly changed by parking the filling at different values. While our previous simulations have been done at $\mu\approx-2$~meV (e.g. Figs.~\ref{realTB}-\ref{T2Rc}), for which $T_{2E}\approx40$~ns, if one changes the full to $\mu\approx-4$~meV the coherence times increases to $T_{2E}\approx70$~ns. Alternatively, changing the filling to $\mu=-0.5$~meV reduces the coherence time to $T_{2E}\approx35$~ns. While the change in coherence is ultimately a result of a reduction in relative spin, this itself stems from the change in spin of the underlying semiconducting states. In particular, at $\mu\approx-4$~meV, the semiconducting band structure hosts two copropagating states with the same spin direction so that difference in spin, inherited by the Andreev states making up the qubit, is relatively small. 

\section{Outlook}
\label{sec:outlook}

Throughout our calculations, we have focused on encoding the ASQ in the lowest energy, in-gap states. Because different states will, in general, have different charge and spin, an encoded qubit can have a qualitatively different dependence of $T_2$ on $\phi$ and $B^z$ which could then be compared with numerical simulations of electric and magnetic noise, analogous to those performed above. In shorter junctions, like those modeled in Sec.~\ref{sec:realistic}, one could encode a qubit in the even occupancy state of the junction\cite{Janvier2015Andreev, hays2018direct, cheung2023photon} while in longer junctions where there are multiple in-gap Andreev doublets, e.g. those found in Sec.~\ref{sec:Andreev states}, one could encode a qubit in any combination of in-gap states\cite{tosi2019spin, hays2020continuous, metzger2020}. 

While our nuclear spin model was sufficient to roughly match experiment, a more sophisticated model could be more predictive. Specifically, using a generalized coupled cluster expansion method\cite{yangAoP20,onizhukPRXQ21}, one could include many, higher-spin defects which would alleviate the need for many  phenomenological parameters introduced to model the noise in Sec.~\ref{sec:Noise Models}. Moreover, such a model could be used to study the regime when energy relaxation is the dominant contributor to decoherence. 

Although we have focused on InAs-inspired model parameters, we could apply our methodology to ASQs realized in other semiconducting materials. In particular, Ge-based qubits have recently received a considerable amount of interest\cite{scappucciNatRevMat21}. Ge spin qubits enjoy an anisotropic $g$-tensor\cite{watzingerNL16}, due to the orbital mixing of the heavy and light holes, and couple to nuclei via a dipole interaction \cite{fischerPRB08,gerardotNAT08}. Because our calculations suggest nuclear spins are the dominant source of decoherence, we expect the decreased sensitivity to nuclei and ability to isotopically purify Ge to enhance Ge-based ASQs coherence. Moreover, because our results are inconsistent with electric fluctuations, we do not expect the strong spin-orbit interaction in Ge to significantly diminish the coherence an ASQ. 

\section{Summary}
\label{sec:summary}
In summary, we have constructed a 1D tight-binding model to study the dependence of the coherence times of ASQs as a function of externally controllable parameters: applied magnetic field, phase difference, and filling of the underlying semiconductor. Because the charge and spin of the Andreev states encoding the ASQ depend on these parameters, the effective coupling to electric and magnetic fluctuators will inherent that dependence. We have matched the qualitative dependence of coherence times in our simulations with those found in experiment~\cite{pita2023direct} and we have identified magnetic fluctuations to be most consistent source of noise in these systems. We predict that by changing the filling of the underlying semiconductor, one could experimentally access ASQs with longer coherence times.  order to further characterize the system.

\section{Acknowledgements}
The authors thank Hugh Churchill, Utkan G\"ung\"ord\"u, and Jamie Kerman for fruitful discussions. The authors are grateful to Marta Pita-Vidal and Arno Bagerbos for helpful comments on their data. This research was funded by the LPS Qubit Collaboratory, and in part under Air Force Contract No. FA8702-15-D-0001. M.H. was supported by an appointment to the Intelligence Community Postdoctoral Research Fellowship Program at the Massachusetts Institute of Technology administered by Oak Ridge Institute for Science and Education (ORISE) through an interagency agreement between the U.S. Department of Energy and the Office of the Director of National Intelligence (ODNI). Any opinions, findings, conclusions or recommendations expressed in this material are those of the authors and do not necessarily reflect the views of the US Air Force or the US Government.

%

\end{document}